\documentclass[a4paper,11pt]{article}
\usepackage{pos}
\usepackage{physics}

\usepackage{color}

\title{$\Lambda$(1405) in the flavor SU(3) limit using a separable potential in the HAL QCD method}

\author*[a,b]{Kotaro Murakami}
\author[c]{Sinya Aoki}

\affiliation[a]{Department of Physics, Institute of Science Tokyo, \\ 
2-12-1 Ookayama, Megro, Tokyo 152-8551, Japan}

\affiliation[b]{Interdisciplinary Theoretical and Mathematical Sciences Program (iTHEMS), RIKEN, Wako 351-0198, Japan}

\affiliation[c]{Center for Gravitational Physics, Yukawa Institute for Theoretical Physics, Kyoto University, \\
Kitashirakawa Oiwakecho, Sakyo-ku, Kyoto 606-8502, Japan}

\emailAdd{kotaro.murakami@yukawa.kyoto-u.ac.jp}
\emailAdd{saoki@yukawa.kyoto-u.ac.jp}
\abstract{
Using the HAL QCD method, 
we investigate S-wave meson-baryon interactions in singlet and two octet channels in the flavor SU(3) limit, where the chiral unitary model predicts that a combination of bound-state poles in these channels corresponds to $\Lambda(1405)$. 
To avoid the singular behavior of the leading-order potentials of these channels in the derivative expansion, we instead employ a separable potential in the time-dependent HAL QCD method. 
To calculate all-to-all propagators in the three-point correlation functions including  $\Lambda$-baryon source operators with zero momentum, we employ the conventional stochastic estimation combined with the covariant approximation averaging. 
Separable potentials both in the singlet and octet channels show attraction without singular behavior.  
Our results of the corresponding phase shifts indicate that the attractive interaction in the singlet channel is stronger than that in the octet. 
Binding energies are consistent with the estimates from the two-point correlation function within one (two) sigma for the singlet (octet) channel, and this ordering of binding energies is consistent with the mass hierarchy suggested by the chiral unitary model.

}
\FullConference{The 41st International Symposium on Lattice Field Theory (LATTICE2024)\\
 28 July - 3 August 2024\\
Liverpool, UK\\}

\begin{document}

\begin{flushright}
RIKEN-iTHEMS-Report-25, YITP-25-11
\end{flushright}

\maketitle
\section{Introduction}
\label{sec:intro}

Properties of hadrons, made of quarks and gluons described by quantum chromodynamics (QCD), play important roles to answer fundamental questions such as what kind of matter can exist in nature.
Especially, understanding the nature of exotic hadrons is one of the essential issues in hadron physics.
Numerical simulations in lattice QCD enable us to study exotic hadrons from the first principles, 
and they become vial tools in this field.
The HAL QCD method~\cite{Ishii:2006ec,Aoki:2009ji,Ishii:2012ssm}, which extracts scattering amplitudes through interaction potentials between hadrons, is one of the methods that can handle exotic hadrons properly even when they are resonances. 
Thanks to the all-to-all calculation technique combined with all-mode averaging~\cite{Akahoshi:2021sxc,Murakami:2022cez}, it is possible to study exotic hadrons in the presence of quark pair annihilations.

In this paper, we investigate $\Lambda(1405)$, an excitation of the $\Lambda$ baryon with $J^{P}=1/2^{-}$, in the flavor SU(3) limit. 
Studies in the chiral unitary model~\cite{Oller:2000fj,Jido:2003cb} suggest that the spectrum corresponding to $\Lambda(1405)$ is explained by a combination of two poles. 
Also, in Refs.~\cite{Jido:2003cb,Bruns:2021krp,Guo:2023wes}, these two poles come from a pole in the singlet channel and one pole in either of two octet channels in the flavor SU(3) limit, with the mass (real part of the pole position) larger in the octet channel than in the singlet. 
Our goal is to see that the meson-baryon interactions generate these poles with the correct mass hierarchy. 
To achieve this goal, we investigate S-wave meson-baryon interactions for the singlet and octet channels in the HAL QCD method. 

In the previous study~\cite{Murakami:2023phq}, we observed that the leading-order (LO) potentials in the derivative expansion have singular behavior in both singlet and octet meson-baryon systems due to zeros of the R-correlators. 
Note, however, that the singular behavior is a technical issue rather than a problem of the HAL QCD method;  potentials with singular behavior can produce correct phase shifts in principle as long as a truncation in the derivative expansion is a good approximation~\cite{Aoki:2021ahj}.
Nevertheless, it turned out to be difficult to calculate reliable binding energies and phase shifts numerically from singular potentials. 
In the current analysis, to avoid such singular behaviors, we introduce separable potentials instead of local potentials in the derivative expansion. 

\section{Time-dependent HAL QCD method with a separable potential}
\label{sec:HAL}
In this section, we explain the time-dependent HAL QCD method~\cite{Ishii:2012ssm} with the usual LO approximation. 
After that, we introduce a separable potential in this method. 
We here consider a two-body hadron system with equal masses for simplicity, but we can apply it to systems with different masses using the semi-relativistic form: replacing the coefficient of the second derivative with $(1+3\delta^2)/8\mu$ where $\delta=(m_1-m_2)/(m_1+m_2)$.

Let us start with an R-correlator defined by 
\begin{eqnarray}
R(\vb{r},t)  
=\frac{\langle H_1(\vb{r+x},t+t_{0})H_2(\vb{x},t+t_{0}) \bar{J}(t_0) \rangle}{C_{H_1}(t)C_{H_2}(t)},
\end{eqnarray}
where $H_1(\vb{x},t)$ and $H_2(\vb{x},t)$ are hadron operators of interest located at $(\vb{x},t)$, $\bar{J}(t_0)$ is the source operator of the two-hadron states at the time $t_0$, and $C_{H_1}(t)$ and  $C_{H_2}(t)$ are two-point correlation functions of each hadron. 
Using the fact that the R-correlator at large $t$ can be expressed by a linear combination of equal-time Nambu-Bethe-Salpeter (NBS) wave functions, we obtain the equation for an interaction potential $U(\vb{r},\vb{r}')$ from the R-correlator at large $t$ as
\begin{eqnarray}\label{eq:timedepeq_general}
\int d^3r' \ U(\vb{r},\vb{r}')R(\vb{r'},t) 
\simeq
\left(\frac{\nabla^2}{2\mu}-\pdv{t} + \frac{1}{8\mu}\pdv[2]{t} \right)R(\vb{r},t),
\end{eqnarray}
where $\mu$ is the reduced mass. 
Hereafter, we assume that $t$ is large enough to satisfy the above equation. 

In the usual calculation, the leading order (LO) approximation in the derivative expansion is given by
\begin{eqnarray}\label{eq:def_LOpot}
U(\vb{r},\vb{r}') \simeq V^{\textrm{LO}}(\vb{r}) \delta^{(3)}(\vb{r}-\vb{r}'),
\end{eqnarray}
from which we obtain
\begin{eqnarray}
V^{\textrm{LO}}(\vb{r}) 
\simeq
\frac{1}{R(\vb{r},t)}\left(\frac{\nabla^2}{2\mu}-\pdv{t} + \frac{1}{8\mu}\pdv[2]{t} \right)R(\vb{r},t). \label{eq:timedepeq_LO}
\end{eqnarray}
Since the R-correlator appears in the denominator in the right-hand side of Eq.~\eqref{eq:timedepeq_LO}, the LO potential has a singular behavior if the R-correlator crosses zero. 
Such a singular behavior was observed in the singlet and octet meson-baryon systems, making it difficult to extract reliable results of observables~\cite{Murakami:2023phq}. 

In the current study, to avoid the singular behavior, we introduce a separable potential as 
\begin{eqnarray}\label{eq:def_sepppot}
U(\vb{r},\vb{r}') \simeq \eta v(\vb{r})v^\ast(\vb{r}')
\end{eqnarray}
instead of the leading-order potential, where $v(\vb{r})$ is a function that depends only on one spacial coordinate and $\eta$ is an overall phase as
$\vert\eta\vert$=1.
We thus obtain 
\begin{eqnarray}\label{eq:timedep_eq_seppot}
\eta v(\vb{r})\int d^3r' \ v^\ast(\vb{r}')R(\vb{r'},t) 
\simeq
\mathcal{D}R(\vb{r},t),
\end{eqnarray}
where 
\begin{eqnarray}\label{eq:def_D}
\mathcal{D}R(\vb{r},t)=\left(\frac{\nabla^2}{2\mu}-\pdv{t} + \frac{1}{8\mu}\pdv[2]{t} \right)R(\vb{r},t).
\end{eqnarray}
We note that $v(\vb{r})$ contains a degree of freedom to choose a constant phase that does not change Eq.~\eqref{eq:def_sepppot},
and we fix the phase so that $\int d^3r' \ v^{\ast}(\vb{r}')R(\vb{r'},t) $ is real and positive in this study.
A extraction of $v(\vb{r})$ and $\eta$ goes as follows, and hereafter we replace ``$\simeq$'' in Eq.~\eqref{eq:timedep_eq_seppot} with ``$=$'' for visibility. 
Multiplying Eq.~\eqref{eq:timedep_eq_seppot} by $R^{\ast}(\vb{r},t)$ and taking an integration over $\vb{r}$, we obtain
\begin{eqnarray}
\eta \left|\int d^3r \ v^{\ast}(\vb{r})R(\vb{r},t)\right|^2 
=
\int d^3r \ R^{\ast}(\vb{r},t)\mathcal{D}R(\vb{r},t).\label{eq:int_timedep_eq_seppot}
\end{eqnarray} 
Since $\vert \eta\vert=1$ and the phase definition of $v(\vb{r})$, we have
\begin{eqnarray}
\int d^3r \ v^{\ast}(\vb{r})R(\vb{r},t) = \sqrt{\left|\int d^3r \ R^{\ast}(\vb{r},t)\mathcal{D}R(\vb{r},t)\right|}, \label{eq:derivation_A}
\end{eqnarray}
which implies
\begin{eqnarray}
&&\eta = \frac{\int d^3r \ R^{\ast}(\vb{r},t)\mathcal{D}R(\vb{r},t)}{\left\vert \int d^3r \ R^{\ast}(\vb{r},t)\mathcal{D}R(\vb{r},t)\right\vert}. \label{eq:derivation_eta} 
\end{eqnarray}
Therefore, using Eq.~\eqref{eq:timedep_eq_seppot}, we finally derive $v(\vb{r})$ as
\begin{eqnarray}
v(\vb{r}) =  \int d^3r \ R(\vb{r},t)\mathcal{D}R^\ast (\vb{r},t)\frac{\mathcal{D}R(\vb{r},t)}{\sqrt{\left|\int d^3r \ R^{\ast}(\vb{r},t)\mathcal{D}R(\vb{r},t)\right|^3}}.\label{eq:derivation_v}
\end{eqnarray}
Since we numerically confirm that $\int d^3r \ R^{\ast}(\vb{r},t)\mathcal{D}R(\vb{r},t)$ is real,
we have $\eta=\pm 1$.
Note that $v(\vb{r})$ does not have singular behavior even if the R-correlator has zero, since the denominator of the right-hand side in Eq.~\eqref{eq:derivation_v} is generally non-zero after the integration over $\vb{r}$.

As seen in Eq,~\eqref{eq:derivation_v}, in order to extract $v(\vb{r})$, we need to take a spacial integration of $R^{\ast}(\vb{r},t)\mathcal{D}R(\vb{r},t)$, which however is given on discrete points on the lattice. 
To reduce systematic errors associated with a summation instead of the integration, we first fit the spacial part of $R^{\ast}(\vb{r},t)\mathcal{D}R(\vb{r},t)$ and then perform the integration over $\vb{r}$ using the fit result. 
To obtain $v(\vb{r})$ for the continuous $\vb{r}$, $\mathcal{D}R(\vb{r},t)$ in Eq.~\eqref{eq:derivation_v} is also fitted.
For fit functions, we employ a sum of Gaussians, $f(\vb{r})=\sum^{N}_{i=1}a_{i}e^{-(r/b_{i})^2}$, with $N=6$ for $R^{\ast}(\vb{r},t)\mathcal{D}R(\vb{r},t)$ and $N=5$ for $\mathcal{D}R(\vb{r},t)$. 

\section{Numerical setups}
\label{sec:setups}
In the present calculation, we use gauge configurations generated with the improved Iwasaki gauge action and the $\order{a}$-improved Wilson quark action at $\beta = 1.83$ and hopping parameters $\kappa_{u}=\kappa_{d}=\kappa_{s}=0.13840$~\cite{Inoue:2011ai}. 
The lattice spacing is $a=0.121(2)~\textrm{fm}$ and the volume is $V=(32a)^4\approx (3.87~\textrm{fm})^4$.
We employ 750 configurations and 32 measurements are performed by shifting the time slice of quark source on each configuration.
We use smeared quark sources in Ref.~\cite{Iritani:2016jie} with $(A,B)=(1.2, 0.2)$. 
The same smearing is employed at the sink with $(A,B)=( 1,{10\over 7})$ to reduce the non-smoothness of Eq.~\eqref{eq:def_D} at short distances~\cite{Akahoshi:2021sxc}. 
Statistical errors of our results are estimated using the jackknife method with 10 jackknife samples.
Masses of the pseudo-scalar meson and positive-parity baryon with octet flavor are extracted from two-point correlation functions of corresponding interpolating operators, and we obtain $m_{8,M}=455.6(1.4)~\textrm{MeV}$ and $m_{8,B}=1163.4(4.1)~\textrm{MeV}$, respectively. 
We also calculate the two-point correlation functions of the $\Lambda$ baryon operator with negative parity, where we find that there is one bound state whose binding energy is $E^{(8)}_{\textrm{bind}}=24.6(15.9)~\textrm{MeV}$ for the octet channel and $E^{(1)}_{\textrm{bind}}=88.6(5.8)~\textrm{MeV}$ for the singlet.

In this study, we focus on the meson-baryon systems in the singlet and the two octet channels, $8_{s}$ and $8_{a}$, where $8_{s}$ ($8_{a}$) is defined to be symmetric (anti-symmetric) by exchanging flavors in meson and baryon. 
In the current analysis, we ignore the coupling between $8_{s}$ and $8_{a}$ channels and perform the single channel analysis in each channel. 
We leave studies on the effect of the channel coupling to future investigations.
We calculate the three-point correlation function using the meson-baryon sink operators in three representations and the singlet $\Lambda$-baryon source operator with negative parity for the singlet channel as well as the octet one for the $8_{s}$ and $8_{a}$ channels. 
In order for this correlation function to have a large overlap with the S-wave ground state, we project the source operator onto zero momentum, where all-to-all propagators appear.
In our calculation of these propagators, we use the conventional stochastic technique combined with the dilutions~\cite{Foley:2005ac} for color/spinor/time components and the s4 dilution for spaces~\cite{Akahoshi:2019klc}. 
We also employ the covariant-approximation averaging~\cite{Shintani:2014vja} combined with the truncated solver method~\cite{Bali:2009hu} using the translational invariance of the baryon operator at the sink. 
Furthermore, to obtain the S-wave component of the potential, we project the sink operators at each spin onto the A$^{+}_{1}$ representation and take an average over spins. 
Note that the potential is independent of the spin component of the baryon since it is related to each other by the rotational symmetry.


\section{Results}
\label{sec:results}

\begin{figure}[t]
    \begin{center}
        \includegraphics[width=0.7\textwidth]{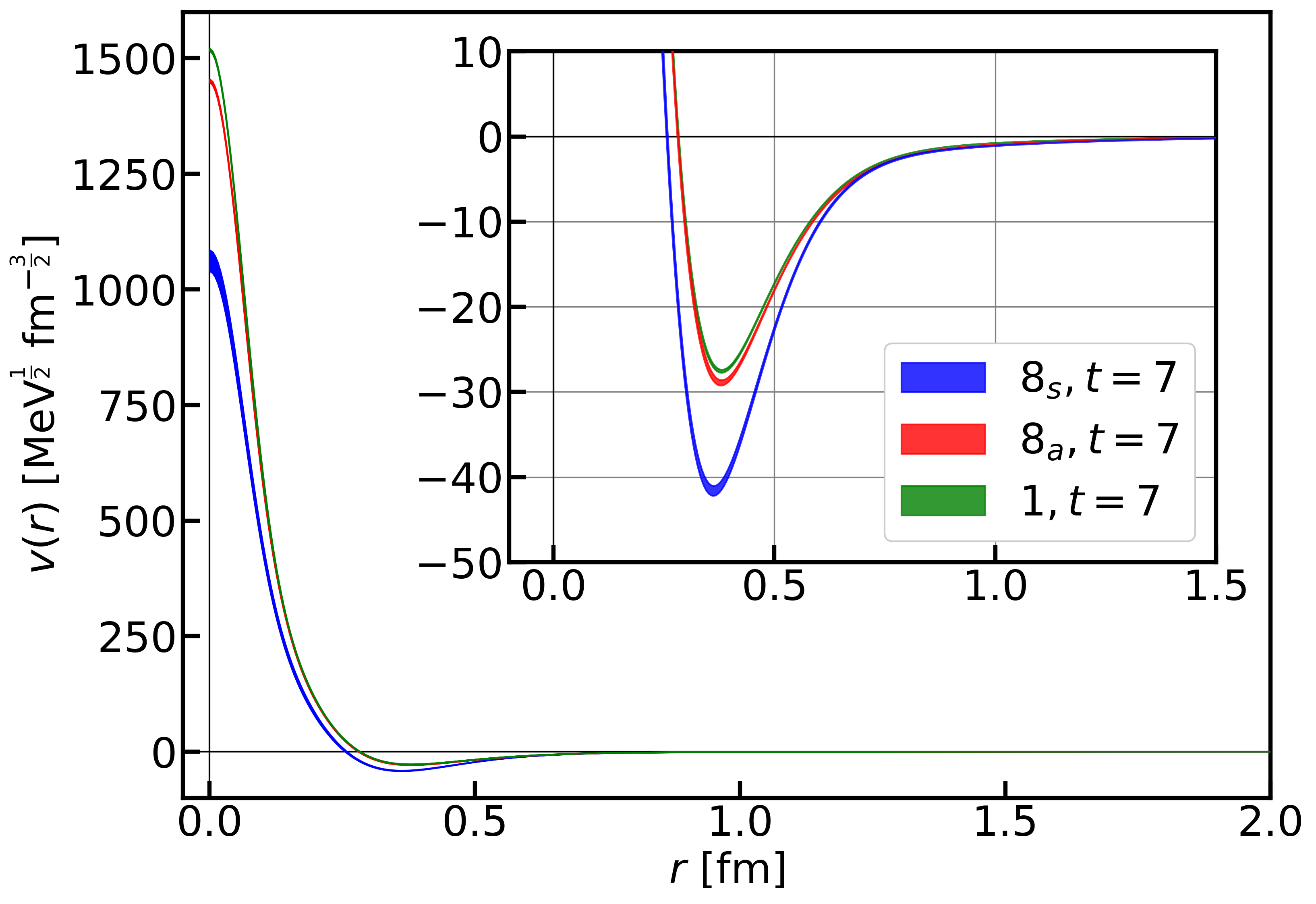}
	    \caption{Separable potentials $v(\vb{r})$ in the $8_s$ (Blue) $8_a$ (Red), and singlet channels (Green) at $t=7$.
        }
	\label{fig:results_seppot}    
    \end{center}
\end{figure}

We calculate Eq.~\eqref{eq:derivation_eta} and  obtain $\eta=-1$ for all three channels, indicating that these meson-baryon systems have attractive potentials. 
We then compute Eq.~\eqref{eq:derivation_v},  which are shown in Figure~\ref{fig:results_seppot}. 
Note that the overall sign of $v(\vb{r})$ is irrelevant since changing the sign does not affect the separable potential itself in Eq.~\eqref{eq:def_sepppot}. 
We find that the shape of the potential is similar between three channels; the potential is dominantly large in the short distance and crosses zero in the middle distance ($r\sim 0.3~\textrm{fm}$). 
Among the three channels, the magnitude in the singlet ($8_s$) channel is the largest (smallest) in the short distance and the smallest (largest) in the middle distance, while there are no significant differences among the three in the long distance.

\begin{figure}[t]
    \begin{center}
        \includegraphics[width=0.7\textwidth]{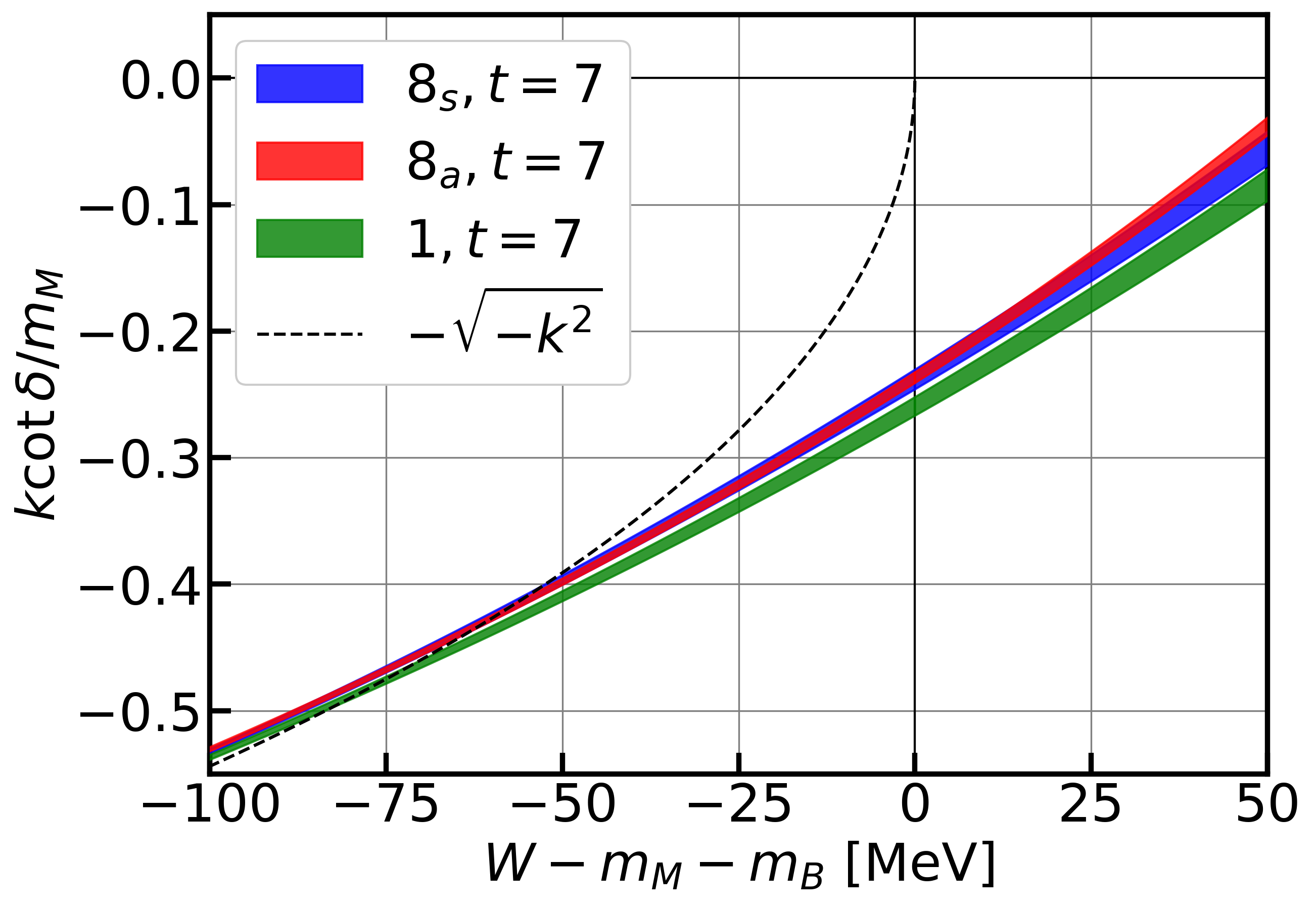}
	    \caption{The $k\cot\delta(k)$ with respect to the total energy from the threshold. 
        Blue, red, and green bands correspond to the results in the $8_s$, $8_a$, and singlet channels, repectively.
        Black dashed line in the negative energy region represents the bound state condition.
        }
	\label{fig:results_kcotd}    
    \end{center}
\end{figure}

We then solve the S-wave Lippmann-Schwinger equation given by
\begin{eqnarray}
t_0(k,k')
=\tilde{U}_{0}(k,k') + \int dq \ \frac{2\mu q^2 \tilde{U}_{0}(k,q)}{k^2-q^2+i\epsilon}t_0(q,k')
\end{eqnarray}
where $t_0(k,k')$ is an S-wave T-matrix and $\tilde{U}_{0}(k,k')$ is a general potential in the momentum space projected onto S-wave. 
For the on-shell momentum $k=k'$, the T-matrix reads $t_0(k,k)=-1/(\pi\mu(k\cot{\delta_0(k)} -ik))$, where $\delta_{0}(k)$ is the S-wave phase shift. 
For a separable potential, the above equation can be reduced to the following relation:
\begin{eqnarray}
k\cot{\delta_{0}(k)}= -\frac{1}{4\pi^2\mu}\frac{\eta}{\tilde{v}(k)^2}\left(1+8\pi\mu\eta \ \mathcal{P}\int dq \ \frac{q^2}{q^2-k^2}\tilde{v}(q)^2\right), \label{eq:kcotd_seppot_single}
\end{eqnarray}
where $\tilde{v}(k)$ is the Fourier transform of $v(\vb{r})$ and $\mathcal{P}\int$ denotes the principal value integral. 
Shown in Figure~\ref{fig:results_kcotd} is our results of $k\cot{\delta_{0}(k)}$ from Eq.~\eqref{eq:kcotd_seppot_single} at $t=7$. 
All results cross the bound state condition, showing the existence of a bound state in each channel. 
Results in the $8_s$ and $8_a$ channels agree with each other within errors in the low-energy regime ($W-m_M-m_B\lesssim 50~\textrm{MeV}$). 
This is in contrast to results of potentials, which clearly show a difference between the two. 
Furthermore, the $k\cot{\delta(k)}$ in the singlet channel is negatively larger than those in $8_s$ and $8_a$ channels, indicating that the interaction in the singlet channel is more attractive than those in two octets. 
These behaviors are qualitatively consistent with those in the SU(3) chiral perturbation theory with the Weinberg-Tomozawa interaction only~\cite{Jido:2003cb}, where two octet channels degenerate and have weaker attractions than the singlet channel.

Finally, to extract the binding energy, we solve the Schr\"{o}dinger equation using the Gaussian expansion method~\cite{Hiyama:2003cu} with the separable potential. 
Our results in $8_s$, $8_a$, and singlet channels are 
\begin{eqnarray}
\begin{aligned}
&E^{(8_s)}_{\textrm{bind}}=57.9(5.5)(^{+5.0}_{-2.9})~\textrm{MeV}, \\
&E^{(8_a)}_{\textrm{bind}}=59.8(3.6)(^{+1.1}_{-11.3})~\textrm{MeV}, \\ 
&E^{(1)}_{\textrm{bind}}=77.0(6.3)(^{+6.8}_{-7.0})~\textrm{MeV},
\end{aligned}\label{eq:BE}
\end{eqnarray}
respectively, where the number in the first (second) parenthesis is the statistical (systematic) error. 
The systematic error is estimated from the difference of results between (1) $t=7$ and $t=6,8$, and 
(2) fit parameters of $R^{\ast}(\vb{r},t)\mathcal{D}R(\vb{r},t)$ and $\mathcal{D}R(\vb{r},t)$  determined using a sum of Gaussians $f(\vb{r})=\sum^{N}_{i=1}a_{i}e^{-(r/b_{i})^2}$
and those including its six mirror images~\cite{Akahoshi:2020ojo} as
\begin{eqnarray}
f_{P}(\vb{r})
= f(\vb{r}) + \sum_{\vb{n}}f(\vb{r}+L\vb{n}),
\end{eqnarray}
where $L$ is the length of the spacial volume and $\vb{n} \in \{ (0,0,\pm1),(0,\pm1,0),(\pm1,0,0) \}$. 
The binding energy in the singlet channel is consistent with the result estimated from the two-point correlation function, $88.6(5.8)~\textrm{MeV}$ within errors.  
On the other hand, the results in  two octet channels are within two sigmas of that from the two-point correlation function $24.6(15.9)~\textrm{MeV}$.
These suggest that our analysis using separable potentials provides more reliable results than the analysis using the LO in the derivative expansion, whose results have large statistical errors due to singular behaviors.  
We also find that the binding energy in the octet is smaller than in the singlet. 
This is also consistent with the prediction by the chiral unitary model~\cite{Jido:2003cb,Bruns:2021krp,Guo:2023wes}.
However, our results in Eq.~\eqref{eq:BE} have sizable statistical and systematic errors, even though
both errors for corresponding separable potentials are rather small.
We observe that binding energies are quite sensitive to short-distance behaviors of separable potentials.
Therefore, a small fluctuation of the potential at short distances causes a large uncertainty of the binding energy.
Furthermore, the result estimated from the two-point correlation function in the octet channel has a large statistical error. 
This may be because a large contamination of scattering states occurs in this two-point function, so that a larger $t$ is needed to extract the binding energy for a shallow bound state.
In order to reduce this problem, an improved technique such as the variational method, should be employed in future studies.


\section{Conclusion}
\label{sec:conclusion}
We investigate the S-wave meson-baryon interactions in the HAL QCD method for singlet and two octet channels in the flavor SU(3) limit, where a combination of poles in these channels corresponds to $\Lambda(1405)$ in the chiral unitary model. 
In the previous analysis, we could not reliably extract observables for these meson-baryon systems since the leading-order potentials in the derivative expansion have singular behaviors due to zeros of  R-correlators.
In this paper, we employ a separable potential instead, where no singular behavior appears because the R-correlator in the denominator is integrated.
In this calculation, we use gauge configurations having the meson and baryon masses $m_{8,M}\approx 456~\textrm{MeV}$ and $m_{8,B}\approx 1163~\textrm{MeV}$, which generate one bound state in each channel. 
In this paper, we perform the single channel analysis for singlet as well as $8_s$ and $8_a$ channels.

The coefficient in the separable potential is found to be $\eta=-1$ for three channels, which shows an attractive interaction. 
Among three channels, separable potentials $v(\vb{r})$ have similar shapes and show some differences in the short and medium distances. 
The $k\cot{\delta(k)}$ in the $8_s$ and $8_a$ channels are almost the same while that in the singlet channel is negatively larger. 
This means that the singlet meson-baryon system is more attractive than two octets, as predicted by the SU(3) chiral perturbation theory.
Binding energies results in $E^{(8_s)}_{\textrm{bind}}=57.9(5.5)(^{+5.0}_{-2.9})~\textrm{MeV}$, $E^{(8_a)}_{\textrm{bind}}=59.8(3.6)(^{+1.1}_{-11.3})~\textrm{MeV}$, and $E^{(1)}_{\textrm{bind}}=77.0(6.3)(^{+6.8}_{-7.0})~\textrm{MeV}$.
These are consistent with the estimate from two-point correlation functions within one sigma in the singlet channel and two sigma in two octets, indicating that our analysis using separable potentials provides reliable binding energies. 
We also observe the hierarchy $E^{(8_s)}_{\textrm{bind}},E^{(8_a)}_{\textrm{bind}}<E^{(1)}_{\textrm{bind}}$, which agrees with  results in the chiral unitary model.

The estimate of the binding energy from the two-point correlation function in the octet channel, however, has a large statistical error, which may be due to a contamination of scattering states. 
An improved technique such as the variational method is needed to reduce these errors, which we leave for future studies. 
We also found that  binding energies are quite sensitive to the short distance behavior of separable potentials, which causes large errors of our results. 
This indicates that the HAL QCD method with a separable potential may enhance the lattice artifact, so that a simulation with a finer lattice is required to study $\Lambda(1405)$ more reliably.
Also, a more general form of the separable potential than a single separable term is necessary to obtain scattering observables in a wider range of energy, especially to extract signals of resonances. 
Such studies, together with coupled channel analysis of the octet channels, are left for future studies.

\acknowledgments
We would like to thank the members of the HAL QCD Collaboration for fruitful discussions.
K.~M. appreciate Drs. D.~Jido, M.~Oka, T.~Sekihara, T.~Hyodo, Y.~Kamiya, A.~Dote, O.~Morimatsu, and T.~Harada for their useful comments. 
We thank Prof. T.~Inoue and ILDG/JLDG~\cite{Amagasa:2015zwb} for providing us with gauge configurations used in this paper. 
We use the lattice QCD code of Bridge++~\cite{Ueda:2014rya} (\url{http://bridge.kek.jp/Lattice-code/}) and its optimized version by Dr. I.~Kanamori~\cite{Kanamori:2018hwh}.
Our research uses computational resources of Wisteria/BDEC-01 Odyssey (the University of Tokyo) provided by the Multidisciplinary Cooperative Research Program in the Center for Computational Sciences, University of Tsukuba, and of supercomputer Fugaku provided by the RIKEN Center for Computational Science through the HPCI System Research Project (Project ID: hp240093). 
K.~M. is supported in part by JST SPRING, Grant Number JPMJSP2110, by Grants-in-Aid for JSPS Fellows (Nos.\ JP22J14889, JP22KJ1870), and by JSPS KAKENHI Grant No.\ 22H04917.
S.~A. is supported in part by JSPS KAKENHI Grant No. 22H00129.
This work is also supported in part by JPMXP1020230411.

\bibliographystyle{JHEP.bst}
\bibliography{Lattice2024_seppot}

\end{document}